\documentclass[compsoc,journal]{IEEEtran}

\usepackage{amssymb,enumitem,subfigure}
\usepackage{soul}
\usepackage{url}
\usepackage{booktabs}       
\usepackage[table]{xcolor}
\usepackage{tcolorbox}
\usepackage[noadjust]{cite}
\usepackage[acronym]{glossaries}

\newcommand{\urllc}{xURLLC}

\newacronym[plural=VUEs,firstplural=vehicular users (VUEs)]{vue}{VUE}{vehicular user}

\newacronym[plural=RSUs,firstplural=roadside units (RSUs)]{rsu}{RSU}{roadside unit}
\newacronym{pdf}{PDF}{probability distribution function}

\newacronym{ar}{AR}{augmented reality}

\newacronym[longplural=base stations]{bs}{BS}{base station}

\newacronym{embb}{eMBB}{enhanced mobile broadband}

\newacronym{jnd}{JND}{just noticeable difference}

\newacronym{mmtc}{mMTC}{massive machine-type communication}

\newacronym{mmw}{mmWave}{milimeter wave}

\newacronym{mbs}{MBS}{macro base station}

\newacronym{lis}{LIS}{large intelligent surfaces}

\newacronym{los}{LoS}{line-of-sight}

\newacronym{ml}{ML}{machine learning}

\newacronym{nlos}{NLoS}{non-LoS}

\newacronym[\glslongpluralkey={radio frequencies}]{rf}{RF}{radio frequency}

\newacronym{rss}{RSS}{received signal strength}

\newacronym{sinr}{SINR}{signal-to-noise ratio}

\newacronym{urllc}{URLLC}{ultra-reliable and low-latency communication}

\newacronym{vr}{VR}{virtual reality}

\newacronym{aoi}{AoI}{age-of-information}

\newacronym{v2v}{V2V}{vehicle-to-vehicle}

\newacronym{tx}{TX}{transmitter}

\newacronym{rx}{RX}{receiver}

\newacronym{gpr}{GPR}{Gaussian process regression}

\newacronym{e2e}{E2E}{end-to-end}

\newacronym{qos}{QoS}{quality-of-service}

\newacronym{qoe}{QoE}{quality-of-experience}

\newacronym{nn}{NN}{neural network}

\newacronym{dnn}{DNN}{deep neural networks}

\newacronym{5g}{5G}{fifth generation}

\newacronym{mf}{MF}{mean-field}

\newacronym{ul}{UL}{upload}

\newacronym{dl}{DL}{download}

\newacronym{lidar}{LiDAR}{light detection and ranging}

\newacronym{fov}{FoV}{field-of-view}

\newacronym{tti}{TTI}{transmission time interval}

\newacronym{harq}{HARQ}{hybrid automatic repeat request}

\newacronym[\glslongpluralkey={radio access technologies}]{rat}{RAT}{radio access technolgy}

\newacronym{noma}{NOMA}{non-orthogonal multiple access}

\newacronym{tsn}{TSN}{time sensitive networking}

\newacronym{sota}{SoTA}{state-of-the-art}

\newacronym{mati}{MATI}{maximum allowable transfer interval}

\newacronym{mad}{MAD}{maximally allowable delay}

\newacronym{sati}{SATI}{stochastic maximum allowable transfer interval}

\newacronym{wncs}{WNCS}{wirelss networked control systems}

\newacronym{admm}{ADMM}{alternating direction method of multipliers}

\newacronym{uav}{UAV}{unmanned aerial vehicle}

\newacronym{hri}{HRI}{human-robot interaction}

\newacronym{rri}{RRI}{robot-robot interaction}

\newacronym{ccdf}{CCDF}{complementary cumulative distribution function}

\newacronym{rgbd}{RGB-D}{color and depth}

\newacronym{ris}{RIS}{reconfigurable intelligent surface}

\newacronym{cococo}{CoCoCo}{communication and control co-design}

\newacronym{prt}{PRT}{perception-reaction time}

\newacronym{rb}{RB}{resource block}

\newacronym{sl}{SL}{split learning}

\newacronym{ee}{EE}{energy efficiency}

\newacronym{snr}{SNR}{signal-to-noise ratio}

\newacronym{irt}{IRT}{isochronous real time}

\newacronym{fl}{FL}{federated learning}

\newacronym{pca}{PCA}{principal component analysis}

\newacronym{bibo}{BIBO}{bounded-input, bounded-output}

\definecolor{aliceblue}{rgb}{0.94, 0.97, 1.0}

\title{\fontsize{21}{22}\selectfont Extreme URLLC: Vision, Challenges, and Key Enablers }

\author{Jihong~Park,
		~Sumudu~Samarakoon,
        ~Hamid~Shiri,
        ~Mohamed~K.~Abdel-Aziz,
        \\$^\dagger$Takayuki~Nishio,
        ~Anis~Elgabli,
        and~Mehdi~Bennis
\thanks{J.~Park, S.~Samarakoon, A.~Elgabli, H. Shiri, M. K. Abdel-Aziz and M.~Bennis are with the Centre for Wireless Communications, University of Oulu, 90014 Oulu, Finland (email: \{jihong.park, sumudu.samarakoon, anis.elgabli, hamid.shiri, mohamed.abdel-aziz, mehdi.bennis\}@oulu.fi). }
\thanks{$^\dagger$T.~Nishio is with the Graduate School of Informatics, Kyoto University, 606-8501 Kyoto, Japan (email: nishio@i.kyoto-u.ac.jp).}
}
\begin{document}

\maketitle

\IEEEpeerreviewmaketitle

\glsresetall

\begin{abstract}
Notwithstanding the significant traction gained by ultra-reliable and low-latency communication (URLLC) in both academia and 3GPP standardization,  fundamentals of URLLC remain elusive.  Meanwhile, new immersive and high-stake control applications with much stricter reliability, latency and scalability requirements are posing unprecedented challenges in terms of system design and algorithmic solutions. This article aspires at providing a fresh and in-depth look into URLLC by first examining the limitations of 5G URLLC, and putting forward key research directions for the next generation of URLLC, coined  eXtreme ultra-reliable and low-latency communication (xURLLC). xURLLC is underpinned by three core concepts: (1) it leverages recent advances in machine learning (ML) for faster and reliable data-driven predictions; (2) it fuses both radio frequency (RF) and non-RF modalities for modeling and combating rare events without sacrificing spectral efficiency; and (3) it underscores the much needed joint communication and control co-design, as opposed to the communication-centric 5G URLLC. The intent of this article is to spearhead beyond-5G/6G mission-critical applications by laying out a holistic vision of xURLLC, its research challenges and enabling technologies, while providing key insights grounded in selected use cases.


\end{abstract}

\glsresetall

\section{Introduction}\label{sec:intro}

\IEEEPARstart{T}{he} overarching goal of \gls{urllc} lies in satisfying the stringent reliability and latency requirements of mission and safety-critical applications.  
Achieving this is tantamount to characterizing statistics of extreme and rare events (e.g., taming the tail of latency distribution), in contrast to the average-based system design~\cite{MehdiURLLC:18,swamy2019monitoring}. 
To remedy to this, 3GPP has been using a brute-force approach centered on  system-level simulations to meet the 99.999\% (5-nine) reliability and 1\,ms latency targets, using a plethora of techniques (short packet transmission, grant-free mechanisms,  leveraging spatial, frequency, and temporal diversity techniques ~\cite{MehdiURLLC:18,ZZZ:mahmood2019time,tech:3gppTR38824}).
While advances have been made in sparse, stationary and  controlled environments with traditional model-based approaches, we still lack a deep understanding of wireless channel dynamics, estimation, stability  of control-loop systems, robustness to unmodeled phenomena, to mention a few.  
These challenges  are further exacerbated  in light of the following recent trends.

\begin{figure}[!t]
	\centering
	\includegraphics[width=\columnwidth]{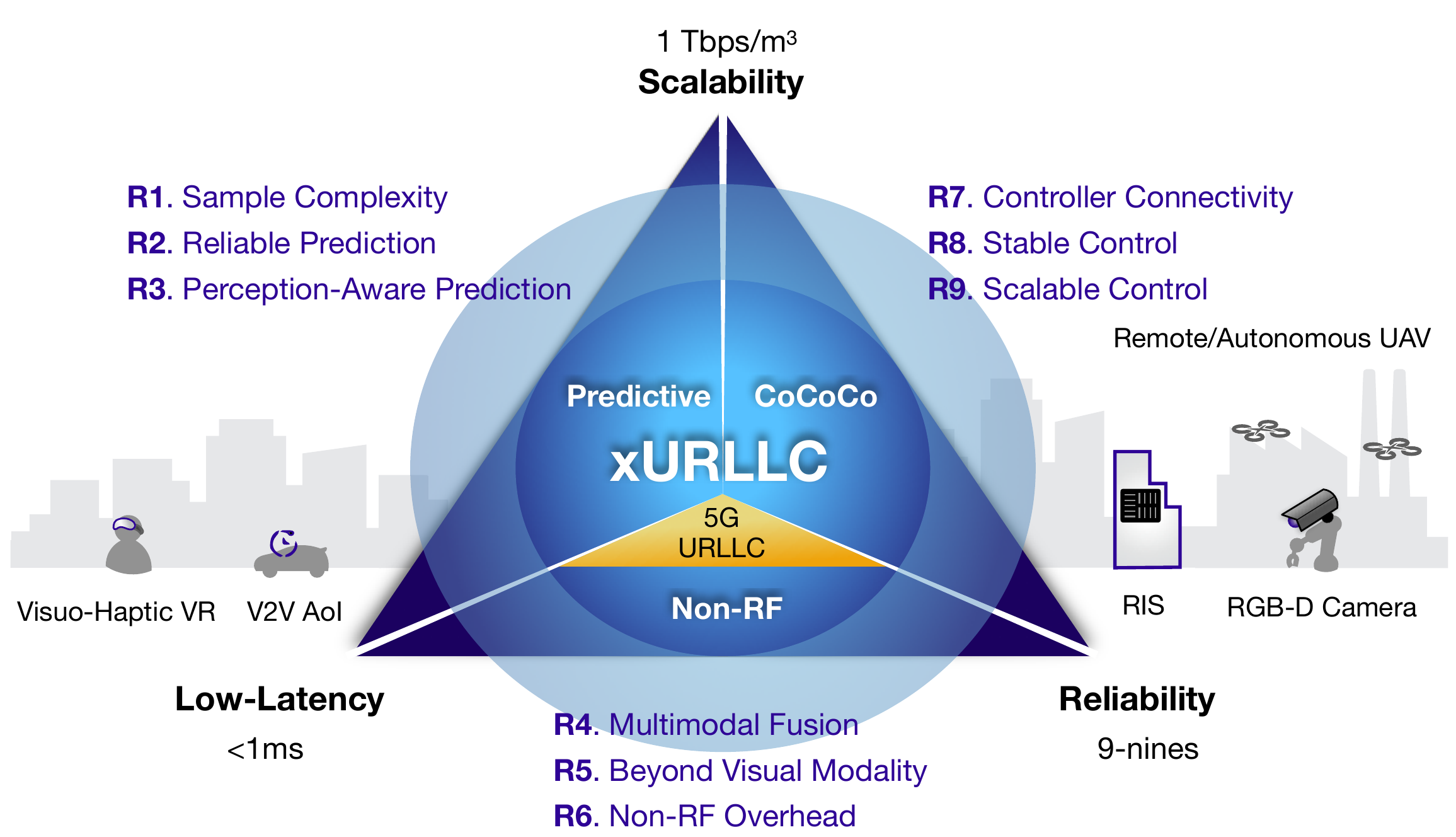} 
	\caption{Anatomy of eXtreme URLLC (\urllc{}): (i) machine learning (ML) based \emph{prediction}, (ii) \emph{non-RF} modality utilization, and (iii) \emph{communication-control co-design (CoCoCo)}.}
	\label{fig:gains}
\end{figure}

On one hand, the emergence of new applications necessitates much \emph{stricter reliability and latency requirements} than those set in 5G \gls{urllc}. 
In particular, high-precision robot control and autonomous vehicles cannot afford 5-nine reliability and millisecond latency~\cite{tech:3gppTR38824}. 
Factory automation over wireless links should guarantee 7-nine reliability and sub-1\,ms latency, similar to those of  the  Ethernet-based  \gls{tsn} and \gls{irt} system standards~\cite{berardinelli2018beyond}.
Meanwhile, the next generation (6G) wireless systems
is advocating 9-nine reliability with 0.1\,ms latency for supporting intelligent systems built upon various perceptual modalities (or Internet of senses) and real-time human-machine interactions~\cite{6genesis2019,ZZZ:mahmood2019time,Saad:Network19}.

On the other hand, \gls{urllc} has become conflated with both \gls{mmtc} and \gls{embb}~\cite{park2018wireless}. 
Unlike the rigid 5G \gls{urllc} design focusing on sparse deployments and short packet transmissions, some applications must simultaneously support massive connections and high data rates, i.e., \emph{scalability}. 
For instance, an autonomous drone swarm in a rescue mission requires not only \gls{urllc} but also {a massive number of wireless control loops} for inter-drone collision avoidance. 
In precision agriculture, vision-based monitoring and remotely controlling sowing robots call for both \gls{urllc} and high-speed data rates. 
In short, the  \gls{embb}-\gls{urllc}-\gls{mmtc} compound is no longer a zero-sum game, {mandating novel solutions to enable scalable support for high data rate mission critical applications.}

As we shall examine, this article discusses the limitations of 5G \gls{urllc}, and puts forward a new research agenda for the next generation of \gls{urllc}, coined \emph{eXtreme \gls{urllc} (\urllc{})}, rooted in three key concepts.

\vspace{5pt}\noindent\textbf{1. Predictive \gls{urllc}.}\quad
5G \gls{urllc} is \emph{reactive} in nature, and is built upon the availability of \emph{known, stationary channel and traffic models}, questioning the adequacy of the  definition of reliability, as noted in~\cite{Popovski:2018aa}. 
In contrast, \urllc{} is essentially \emph{predictive}, leveraging the recent advancement in \emph{\gls{ml}} to enable highly accurate predictions of channels, traffic, states, and other key performance indicators~\cite{park2018wireless}. 
A fundamental research question addressed by predictive \gls{urllc} is summarized as follows.

\vspace{3pt}
\begin{tcolorbox}[colback=aliceblue, boxrule=0pt, boxsep=3pt,left=1.5pt,right=1.5pt,top=2pt,bottom=2pt, sharp corners]
\begin{itemize}
\item[\textbf{Q1.}] {Can wireless environments (channels, interference, services, etc.) be reliably predicted based on past data samples? and under what future horizon?}
\end{itemize}
\end{tcolorbox}

\noindent The challenges raised by \textbf{Q1} and associated research agenda~({\textbf{R1}-\textbf{3}}) are discussed in Sect.~\ref{sec:predictive_urllc}, followed by selected use cases.

\vspace{5pt}\noindent\textbf{2. Non-\gls{rf} Aided \gls{urllc}.}\quad
By design 5G \gls{urllc} is \emph{\gls{rf}}-based  and requires investing wireless resources for channel probing and estimation. 
By contrast, \urllc{} exploits \emph{non-\gls{rf} modalities}, such as \gls{rgbd} images for channel prediction \cite{Koda:2019aa}.
This data  provides rich features for predicting extreme and sudden events (e.g., blockages), which cannot be done with  current \gls{rf}-based solutions, due to lack of statistical relevance and/or expensive acquisition under limited radio resources. 
By extension, one can utilize \glspl{ris} and metasurfaces~\cite{Saad:Network19} to tune channel randomness by manipulating surface reflections, while inducing higher \gls{ee}. 
Enjoying these benefits hinges on addressing the following question.

\vspace{3pt}
\begin{tcolorbox}[colback=aliceblue, boxrule=0pt, boxsep=3pt,left=1.5pt,right=1.5pt,top=2pt,bottom=2pt, sharp corners]
\begin{itemize}
\item[\textbf{Q2.}] {How to transfer and fuse non-\gls{rf} and \gls{rf} modalities with minimum overhead to enable \textsf{\urllc{}}?}
\end{itemize}
\end{tcolorbox}

\noindent Sect.~\ref{sec:non_rf} addresses the challenges and opportunities ({\textbf{R4}-\textbf{6}}) raised by \textbf{Q2} through the lens of exemplary use cases.

\vspace{5pt}\noindent\textbf{3. Control Co-Designed \gls{urllc}.}\quad
In 3GPP parlance, communication reliability is calculated by counting erroneous packets divided by the total transmitted packets during an observed time period \cite{tech:3gppTR38824}. 
In contrast, \urllc{} cares about whether \emph{consecutive packet errors or losses} disrupt the control operation. 
Understanding control dynamics provides a natural (yet untapped) opportunity to relax the very stringent latency and reliability requirements, making \emph{\gls{cococo}} a core concept in \urllc{}. 
To reach this goal, one should take into account wireless channel dynamics in control systems, through which the received state observations and actuating commands may be outdated and distorted. 
Moreover, relaxing communication latency and reliability requirements, while guaranteeing control stability and safety against external perturbations, internal state fluctuations, and inter-agent collision is of paramount importance, raising the following question.

\vspace{3pt}
\begin{tcolorbox}[colback=aliceblue, boxrule=0pt, boxsep=3pt,left=1.5pt,right=1.5pt,top=2pt,bottom=2pt, sharp corners]
\begin{itemize}
\item[\textbf{Q3.}] {Can \gls{urllc} requirements be relaxed by taking into account control dynamics, while ensuring control stability? }
\end{itemize}
\end{tcolorbox}

\noindent In Sect.~\ref{sec:control_commun}, \textbf{R7}-\textbf{R9} discuss the challenges and opportunities raised in \textbf{Q3}  through selected case studies, followed by conclusions in Sect.~\ref{sec:conclusions}.

\section{Predictive URLLC}
\label{sec:predictive_urllc}

5G \gls{urllc} focuses on characterizing extreme events at the cost of spectral efficiency, limiting its scalability. 
In doing so, 5G \gls{urllc} presumes a static channel model that fails to capture non-stationary channel dynamics and exogenous uncertainties (e.g., out-of-distribution or other under-modeled rare events), which are common in uncontrolled environments. In contrast, \urllc{} aims at proactive decision making powered by \gls{ml}, in which proactiveness offers available resources to satisfy 9-nine reliability within 0.1\,ms latency, which is on par with  Ethernet-based \gls{tsn} and \gls{irt}~\cite{berardinelli2018beyond}. 
Furthermore, in contrast to the static and model-based paradigm, {\urllc{} allows to communicate by learning from data samples even under non-stationary and unpredictable environments.} 
Examples include latency estimation, 
\gls{aoi} \cite{Khairy:CL20GPR}, and  traffic demand prediction~\cite{ParkGC:18} based on users' visuo-haptic perceptions.

\subsection{Challenges and Opportunities}
The adoption of \gls{ml} entails novel challenges and research opportunities in \urllc{}, as we shall examine.

\vspace{5pt}\noindent\textbf{R1. {Sample Complexity.}}\quad 
Making predictions using an \gls{ml} model, i.e., inference, should be preceded by model training. 
A trained model is valid so long as the training data distribution is unchanged; otherwise, the outdated model must be re-trained. 
The interval of this \emph{continual learning}, i.e., training convergence time, should be sufficiently small compared to the temporal channel evolution dynamics. 
Training convergence analysis quantifies the required number of training samples to achieve a target accuracy, i.e., \emph{sample complexity}~\cite{park2018wireless}. 
Due to the lack of samples at a single location, taming sample complexity requires communication. 
\Gls{fl} addresses this problem by periodically exchanging locally trained model parameters, rather than instantly exchanging raw samples~\cite{park2018wireless},
thereby reducing the cost of predictive \gls{urllc}.

\vspace{5pt}\noindent\textbf{R2. Reliable Prediction.}\quad 
Predictive \gls{urllc} improves communication reliability, as long as the \gls{ml} prediction is reliable. 
Although traditional deep \gls{nn} models can achieve high prediction accuracy, they do not report the reliability of their prediction. 
To measure reliability against unseen training samples, one needs to quantify the \emph{generalization error}, defined as the difference between the expected loss across the entire dataset and the empirical training loss~\cite{park2018wireless}. 
Another solution is to leverage Bayesian learning methods such as \emph{\gls{gpr}} that provide the prediction confidence via the variance of the posterior distribution~\cite{Khairy:CL20GPR}. 
Last but not least, \emph{adversarial training} improves reliability against non-stationary data distributions due to time-varying channels, malfunctions and attacks, by training with synthetic adversarial data samples.

\vspace{5pt}\noindent\textbf{R3. Perception-Aware Prediction.}\quad  Look-ahead forecasting offers more available resources, at the expense of accuracy. 
Prediction horizon should therefore be minimized by utilizing perceptual characteristics. 
For driving scenarios, the prediction horizon can be determined by the driver's \emph{\gls{prt}} that is around 2.5\,s for human drivers and several milliseconds for driverless cars. 
When human-driving and driverless cars coexist, the \gls{prt} of both driving agents may increase if they do not understand the other agents' reasoning. 
The \emph{\gls{hri}} should therefore be closely investigated. 
In high-precision control applications involving multiple perceptual modalities (e.g., vision, touch, Lidar, etc.), the perceptual relationships and their integrated  resolutions measured using the \emph{\gls{jnd}}  ought to be considered~\cite{ParkGC:18}.


\vspace{3pt}

The following subsections discuss some of the issues raised in \textbf{R1}-\textbf{R3} in \gls{v2v} and \gls{vr}/\gls{ar} scenarios.


\begin{figure}[!t]
	\centering
		\includegraphics[width=\linewidth]{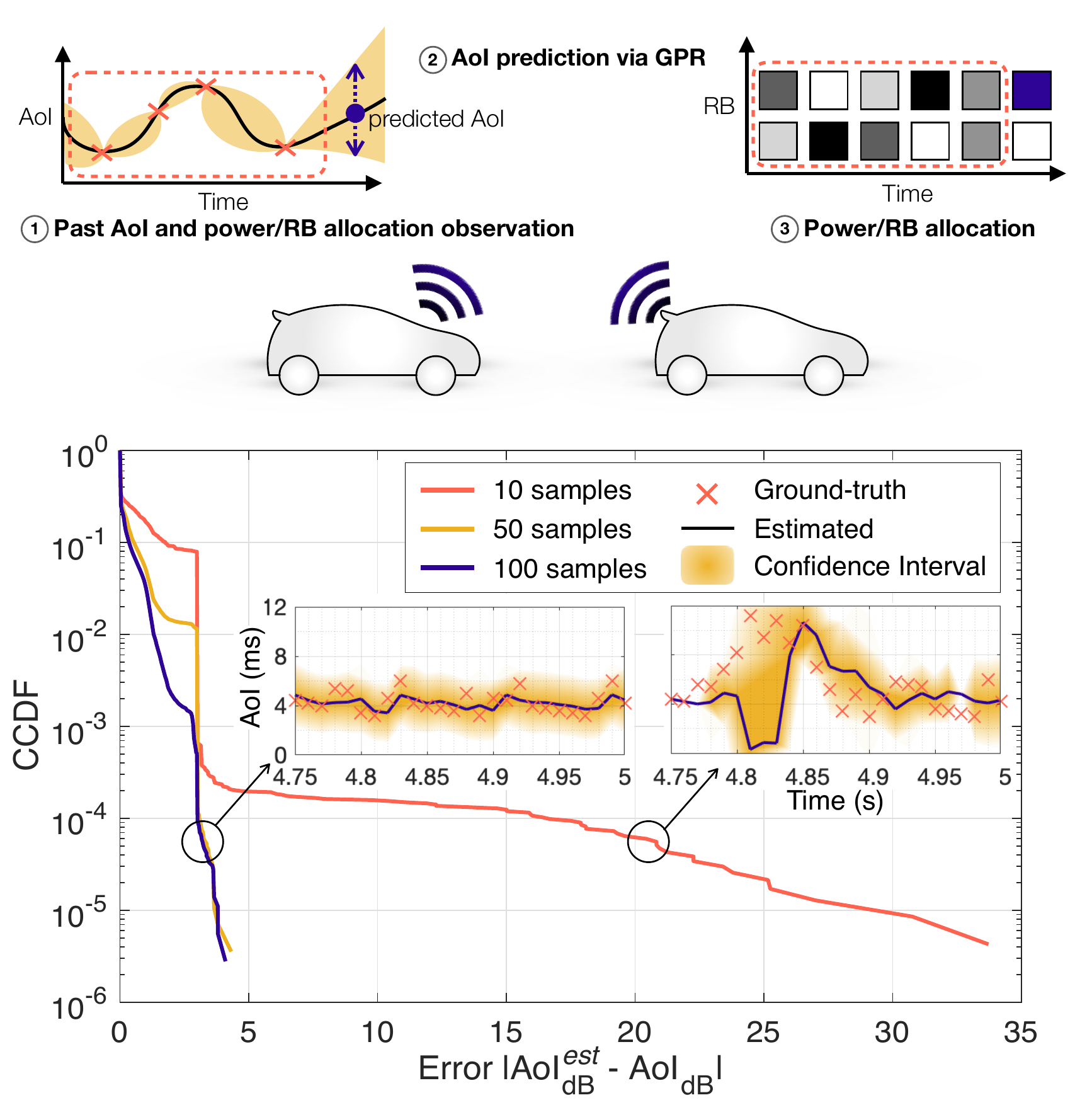} 
	\caption{Tail error distribution between the actual \gls{aoi} and its estimated value via \gls{gpr}, in a \gls{v2v} communication scenario.}
	\label{fig:use_case_ml}
\end{figure}

\begin{figure}[!t]
	\centering
		\includegraphics[width=\columnwidth]{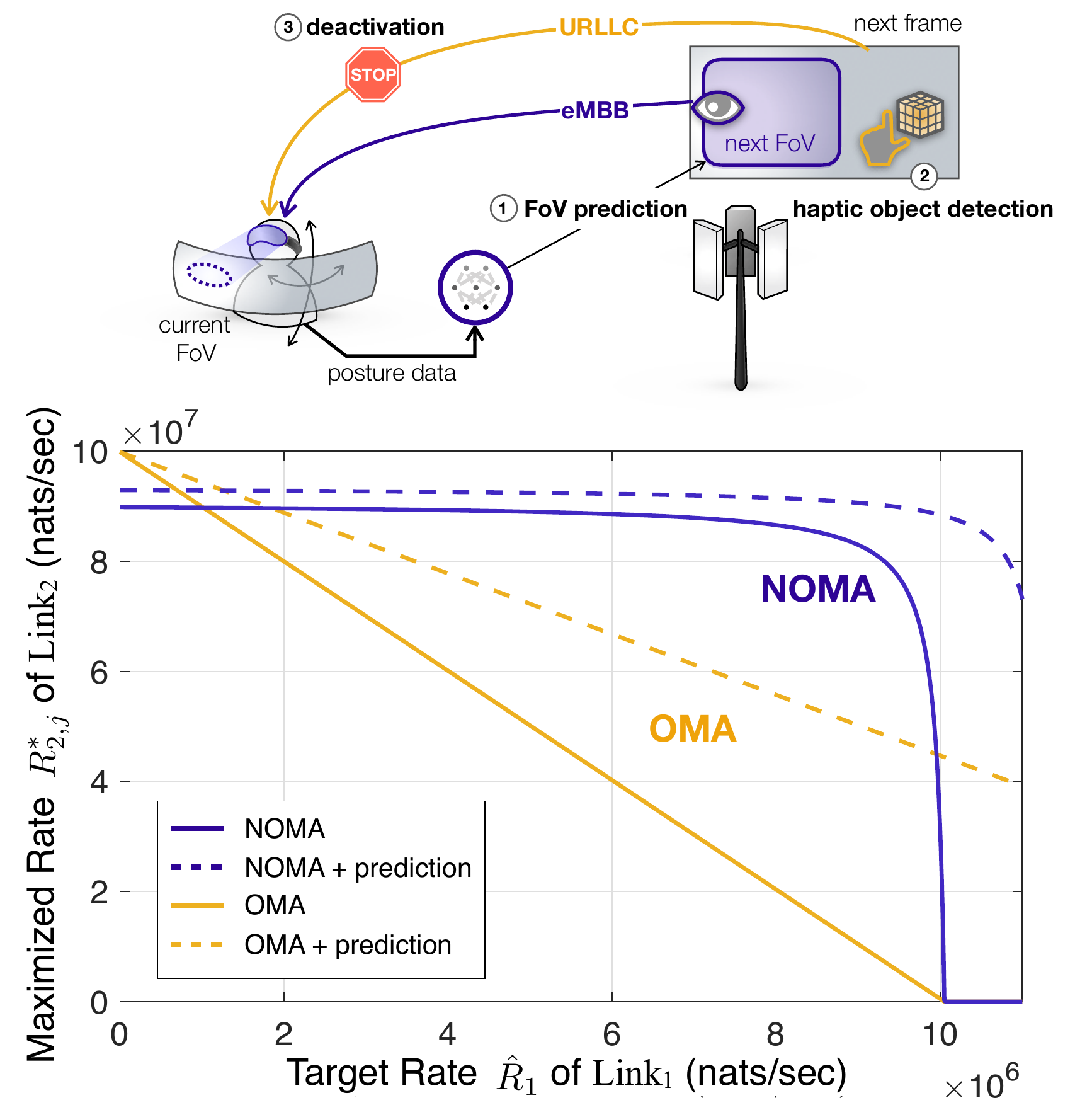} 
	
	\caption{\gls{urllc}-\gls{embb} rate region when \gls{urllc} resource is constantly reserved (solid curves) or is dynamically allocated via visuo-haptic \gls{vr} traffic demand prediction (dotted curves).
}
	\label{fig:use_case_perception}
\end{figure}

\subsection{Use Cases}

\subsubsection{Predictive \glsentryshort{aoi} for Ultra-Reliable \glsentryshort{v2v} Communication}

Ensuring the freshness of safety messages is crucial in \gls{v2v} communication, which is measured using the notion of \gls{aoi}. 
\Gls{aoi} is defined as the time duration from the message generation to reception \cite{Khairy:CL20GPR}. 
Estimating future \gls{aoi} is however a daunting task, as it depends on past resource allocation decisions and channel dynamics. 
To overcome this difficulty, \gls{gpr} can be utilized as follows. 

\vspace{5pt}\noindent\textbf{Scenario.}\quad 
There are $20$  transmitting-receiving vehicle pairs driving in a Manhattan grid scenario. 
Under this time-varying \gls{v2v} channel, every transmitter locally determines its transmission power and \gls{rb} selection, such that the receiver's \gls{aoi} is bounded by a predefined threshold with a target reliability. 
To this end, each transmitter runs \gls{gpr} by feeding its past \gls{aoi} and \gls{rb} selections, yielding the next transmission power and \gls{rb} selection. 

\vspace{5pt}\noindent\textbf{Results.}\quad 
Fig.~\ref{fig:use_case_ml} plots the tail distribution of the error between the true and estimated \gls{aoi} for a given power and \gls{rb} decision. 
The more samples are used, the sharper the tail distribution is (see \textbf{R1}). 
Furthermore, the prediction reliability is fully characterized by the posterior distribution's variance, which decreases with more samples (see~\textbf{R3}).


\subsubsection{\glsentryshort{vr}/\glsentryshort{ar} Perception-Aware Proactive Network Slicing}

Supporting multimodal perceptions at a single device is becoming increasingly important in 5G and beyond. 
Indeed, a mobile \gls{vr} user may simultaneously look at and touch a virtual object. 
Such concurrent visual and haptic perceptions should be synchronously received at the user as experienced in real life, so as to avoid cybersickness while increasing  immersion in virtual spaces. 
Since these visual and haptic modalities have distinct rate and latency requirements, they should be supported through separate \gls{embb} and \gls{urllc} links, while mitigating their inter-modal interference. 
By utilizing their perceptual relationship \gls{embb}-\gls{urllc} links can be proactively sliced as follows.

\vspace{5pt}\noindent\textbf{Scenario.}\quad
Using \gls{embb}-\gls{urllc} links, a \gls{bs} serves a downlink \gls{vr} user watching a visuo-haptic interactive movie, while ensuring a target perceptual resolution (see~\textbf{R3}). 
\Gls{jnd} quantifies the perceptual resolution (e.g., 3\,mm minimum detectable object size), which is given by the harmonic mean of the individual links' packet error rates~\cite{ParkGC:18}. 
Unfortunately, satisfying the multimodal perception requirement consumes huge wireless resources. 
To resolve this problem, we utilize the fact that haptic experiences are limited by touchable objects within the visual \gls{fov}. 
Consequently, using a recurrent \gls{nn} and feeding  past~\glspl{fov}, the \gls{bs} deactivates the \gls{urllc} link, if there exists no touchable object within the future \gls{fov}.

\vspace{5pt}\noindent\textbf{Results.}\quad 
Fig.~\ref{fig:use_case_perception} shows that proactive \gls{urllc} deactivation improves the \gls{embb} data rate under both orthogonal and non-orthogonal slicing methods, particularly for a high \gls{urllc} target data rate, i.e., supporting high-resolution haptic experiences. 
The prediction horizon of \gls{fov} was 5~frames ahead, corresponding to 41.7\,ms under 120\,Hz frame~rate.

\section{Non-RF Aided URLLC}
\label{sec:non_rf}

Spurred by recent advances in \gls{ml} and computer vision,  leveraging non-\gls{rf} modalities (e.g., \gls{rgbd} and cameras, Lidar, etc.) is crucial for confronting the extreme event prediction problem, without sacrificing spectral efficiency. 
Compelling non-\gls{rf} aided \gls{urllc} use cases include vision-based channel prediction and mobility management, vision-aided coordination and control of robotic swarms.
To overcome the issue of occlusion,  transferring visual modalities into \gls{rf} expedites channel blockage predictions without  pilot signaling \cite{Koda:2019aa},
allows high-precision location prediction and tracking \cite{Alahi:2015aa},  to mention a  few.  
Not only that, for enabling scalable \urllc{},  non-\gls{rf} modalities provide yet another source of diversity-enhancements, free from \gls{rf} resource constraints and negligible signaling overhead.


\subsection{Challenges and Opportunities}

Smartly incorporating non-\gls{rf} modalities is crucial for  enabling \urllc{} with negligible overhead. 
This rests on addressing the following challenges and research opportunities.

\vspace{5pt}\noindent\textbf{R4. Multimodal Fusion.}\quad
Different types of data have distinct spatio-temporal resolutions (e.g., image sizes, frame rates, and sensing/sampling rates) and their appropriate processing methods (e.g., convolutional \gls{nn} for vision). 
Efficiently fusing multiple modalities while balancing their useful features is a critical challenge. 
\emph{\Gls{sl}} is a powerful framework, in which an \gls{nn} consists of a shared upper segment connected to multiple lower segments fed by different types of data. 
\Gls{sl} can effectively fuse multiple modalities by applying different data-specific architectures to the lower segment, while balancing the aggregating weights at the upper segment~\cite{Koda:2019aa}.

\vspace{5pt}\noindent\textbf{R5. Beyond Visual Modality.}\quad
Besides vision, there exist other non-\gls{rf} modalities to enable \urllc{}. 
\emph{Accelerometer} information is one possible candidate, from which mobility patterns can be extracted, thereby estimating blockage duration. 
Another possibility is \emph{\gls{ris}}-endowed walls based on the idea of manipulating signal amplitude, phase, reflection angle, and polarization via \gls{ris} such that the desired signals and interference can be engineered. 
%
%
To effectively fuse these non-\gls{rf} modalities, their pros and cons should be carefully examined vis-a-vis  \gls{rf}-based \gls{urllc}.

\vspace{5pt}\noindent\textbf{R6. Non-\gls{rf} Overhead.}\quad
Utilizing non-\gls{rf} data does not come for free, and its extra energy cost for acquisition, processing, and control cannot be neglected.
Indeed, to enable vision-aided channel estimation, training images need to be collected from multiple cameras to overcome each camera's limited FoV, consuming communication energy. 
Extracting hidden useful features via \gls{pca} or convolutional \gls{nn} entails processing energy. 
Balancing vision and \gls{rf} modalities in the fusion operations requires an optimization procedure that consumes additional computing energy. These energy footprints may negate the effectiveness of non-\gls{rf} modalities, notably compared to advanced digital beamforming in 6G~\cite{6genesis2019,Saad:Network19}. 
Therefore, \emph{\gls{ee}}, defined as the ratio of the performance gain to the total energy consumption, should be carefully examined.

\vspace{5pt}

The following subsections tackle  \textbf{R4}-\textbf{R6} focusing on \gls{rgbd} image based \gls{mmw} channel prediction and \gls{ml} based  energy-efficient \gls{ris} use cases.



%

\begin{figure}[t]
    \centering
    	\includegraphics[width=\columnwidth]{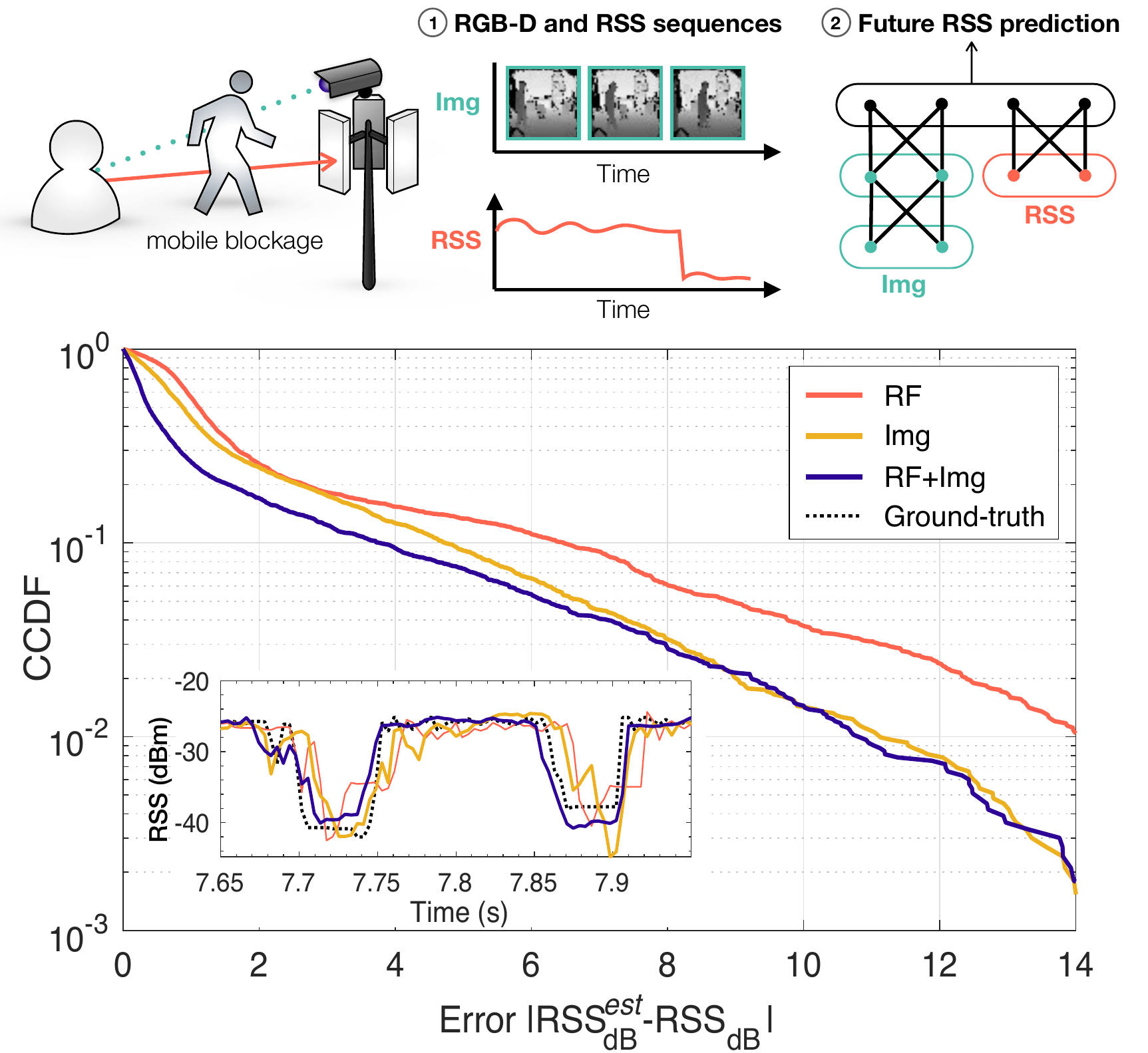} 
    \caption{Tail distribution of the error between actual \gls{mmw} \gls{rss} and its estimated value by fusing past \gls{rss} data and \gls{rgbd} images.
    }
    \label{fig:use_case_non_rf}
\end{figure}

\subsection{Use Cases}

\subsubsection{\glsentryshort{rgbd} Aided \glsentryshort{mmw} Received Power Prediction}

\gls{rf} signals do not always have sufficient features for high-accuracy prediction. 
Predicting future \gls{mmw} channels is one example, in which  predictions using past \gls{rf} signals fail to detect sudden transitions between \gls{los} and \gls{nlos} conditions due to pedestrian blockages~\cite{Koda:2019aa}.
This is where visual modalities comes to the rescue, namely where a sequence of camera images containing sufficient features to predict channel blockages complements to the \gls{rf} modality, as detailed next.

\vspace{5pt}\noindent\textbf{Scenario.}\quad
Consider an \gls{rgbd} camera with 30\,Hz frame rate observing a 60\,GHz \gls{mmw} channel that is randomly blocked by two moving pedestrians.
Our goal is to predict 120\,ms ahead the received power using past camera images and received powers observed at the same time.
To this end, a split \gls{nn} is considered (see \textbf{R4}), comprising: 
1) two convolutional layers that extract features from images; 
2) another single-layer \gls{nn} whose output dimension is the same as the output dimension of 1); 
and 3) a recurrent \gls{nn} layer that concatenates the outputs of 1) and 2) into a sequence as its input,  thereby performing a time-series prediction of the future \gls{mmw} received power.

\vspace{5pt}\noindent\textbf{Results.}\quad
Fig.~\ref{fig:use_case_non_rf} shows that the received power prediction using both images and \gls{rf} signals (\textsf{\gls{rf}+Img}) achieves the highest accuracy while precisely detecting the \gls{los}/\gls{nlos} transitions. 
By contrast, the baseline predictions using either \gls{rf} signals (\textsf{\gls{rf}}) or images (\textsf{Img}) fail to accurately predict the transitions or short-term channel fluctuations for a given \gls{los} or \gls{nlos} condition. 
While \textsf{\gls{rf}+Img} is effective in minimizing the mean prediction error, the tail probability shows that extremely large error occurrences are minimized under \textsf{Img}, {calling for further optimizing the multimodal fusion using a tail-risk minimization framework~\cite{MehdiURLLC:18}.}


\subsubsection{ML Based Energy-Efficient \glsentryshort{ris}} \label{Sec:Case_RIS}
Ensuring reliable connectivity with extremely low energy consumption is instrumental in realizing scalable \urllc{}. 
Towards achieving this goal, \Gls{ris} is a promising enabler, in which a large number of low-cost reflectors within a planar array passively shift the phases of incident \gls{rf} signals (see~\textbf{R5}).  
This begs the question of \textit{how to design a low-complexity \gls{ris} controller with minimal signaling overhead, while achieving high \gls{ee}.}

\vspace{5pt}\noindent\textbf{Scenario.}\quad
An \gls{ris} with $64$ elements serves a single user, by reflecting the signals transmitted from a single \gls{bs}. 
The entire elements are equally divided and controlled by $K$~controllers, each of which is a fully-connected $3$-layer \gls{nn} with $N$ neurons per layer. 
By feeding in the user location, the \gls{nn} outputs either $0$ or $\pi$ phase shift per element. 
The \gls{nn} is trained via supervised learning using $650$ samples, by minimizing the \gls{snr} difference between the proposed method and the ground truth found via exhaustive search. 
Subsequently, for a given new user's location,  the \gls{ris} phase shift is inferred using the trained \gls{nn} model.

\begin{figure}[t]
    \centering
    	\includegraphics[width=\columnwidth]{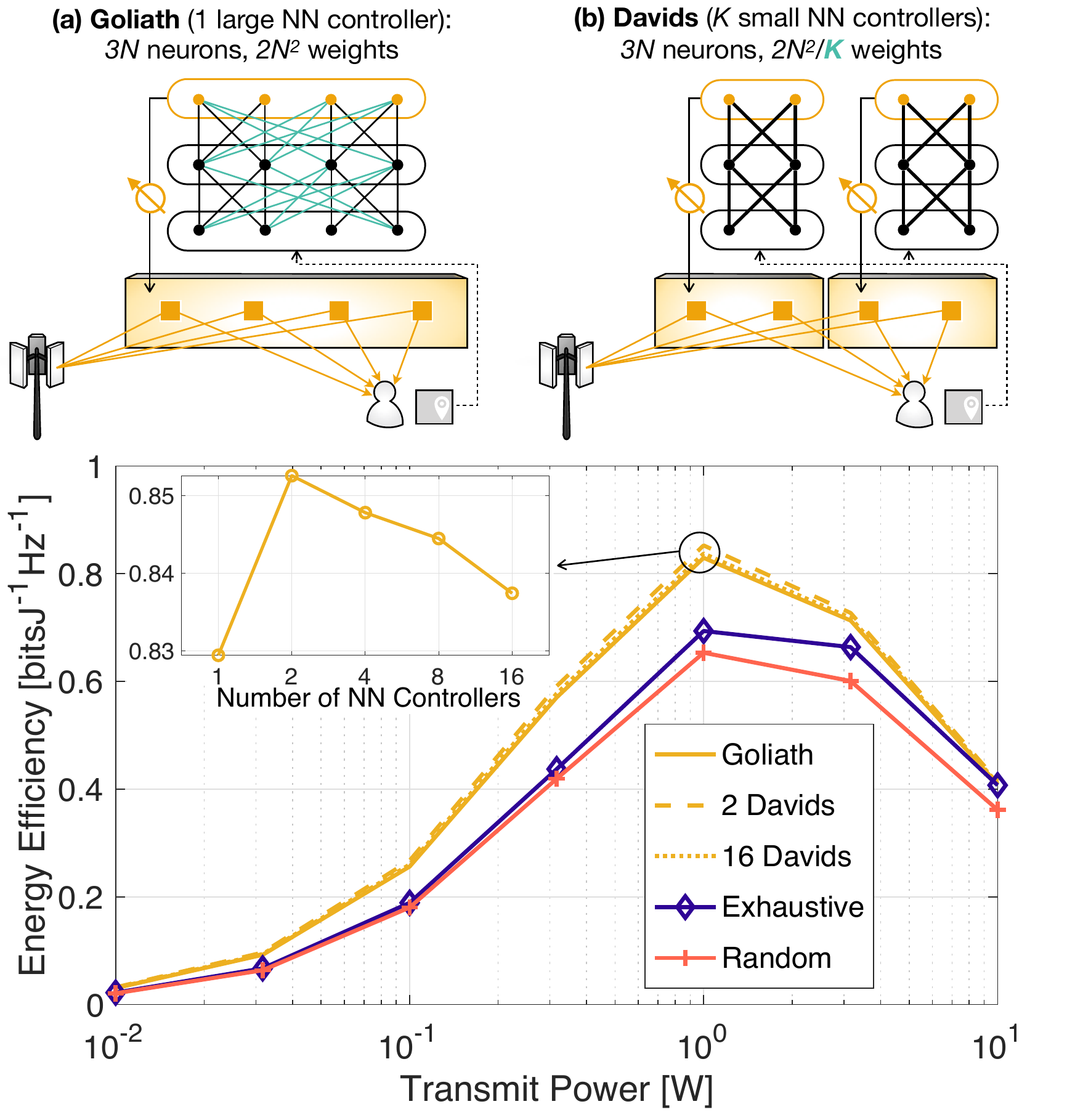} 
    \caption{\Gls{ris} system \gls{ee} under a single large \gls{nn} controller (\textsf{Goliath}, left) or multiple small \gls{nn} controllers (\textsf{Davids}, right), as the \gls{bs} transmission power increases.}
    \label{fig:use_case_ris}
\end{figure}

\vspace{5pt}\noindent\textbf{Results.}\quad
Fig.~\ref{fig:use_case_ris} shows that the proposed method achieves higher \gls{ee} (see~\textbf{R6}), defined as spectral efficiency per total energy consumption (excluding the \gls{bs} transmission power), than a random phase shifting baseline. 
Compared to  exhaustive search, the proposed method yields almost the same spectral efficiency without dissipating energy as done in exhaustive search, resulting in higher \gls{ee}. 
To further improve \gls{ee}, increasing the \gls{bs} transmission power is shown to be effective up to a certain inflection point after which the bottleneck stems from the binary phase shifting, calling for the controller's design and optimization.
Compared to a single large \gls{nn} controller (\textsf{Goliath}), $K$ small \gls{nn} controllers (\textsf{Davids}) consume less energy that is proportional to the number of weights ($2N^2\!/K$). 
On the contrary, too many \textsf{Davids} incur much fewer weights, reducing the \gls{nn} model capacity. 
Under this energy consumption and model capacity trade-off, \gls{ee} is maximized at $K=2$.




\section{Control Co-Designed URLLC}
\label{sec:control_commun}

Real-time control over wireless links is a cornerstone application in \gls{urllc}, requiring the strictest reliability and latency targets \cite{MehdiURLLC:18,Saad:Network19,eisen2019control}. 
Conversely, control is a domain where relaxing the \gls{urllc} requirements can be maximized, {thereby enabling scalability}. 
This hinges on identifying the importance of each transmission packet in control operations subject to, for example, the \gls{mati}, the \gls{mad}, and \gls{aoi}~\cite{Khairy:CL20GPR}. 
This system design is at odds with 5G \gls{urllc} focusing solely on over-the-air and one-way transmission errors with equal importance for all packets. 
Not only that, a key requirement overlooked in 5G \gls{urllc} is stability, which makes \gls{cococo} of utmost importance for guaranteeing physical stability \cite{Shiri:2019aa}. 
Indeed, once a device drifts away from controllable states, further communication becomes useless and wasteful. 
\urllc{} should therefore play a pivotal role in, for instance, avoiding collisions of autonomous vehicles and guaranteeing to reach  a target destination.

\begin{figure*}[!ht]
	\centering
	\includegraphics[width=.98\textwidth]{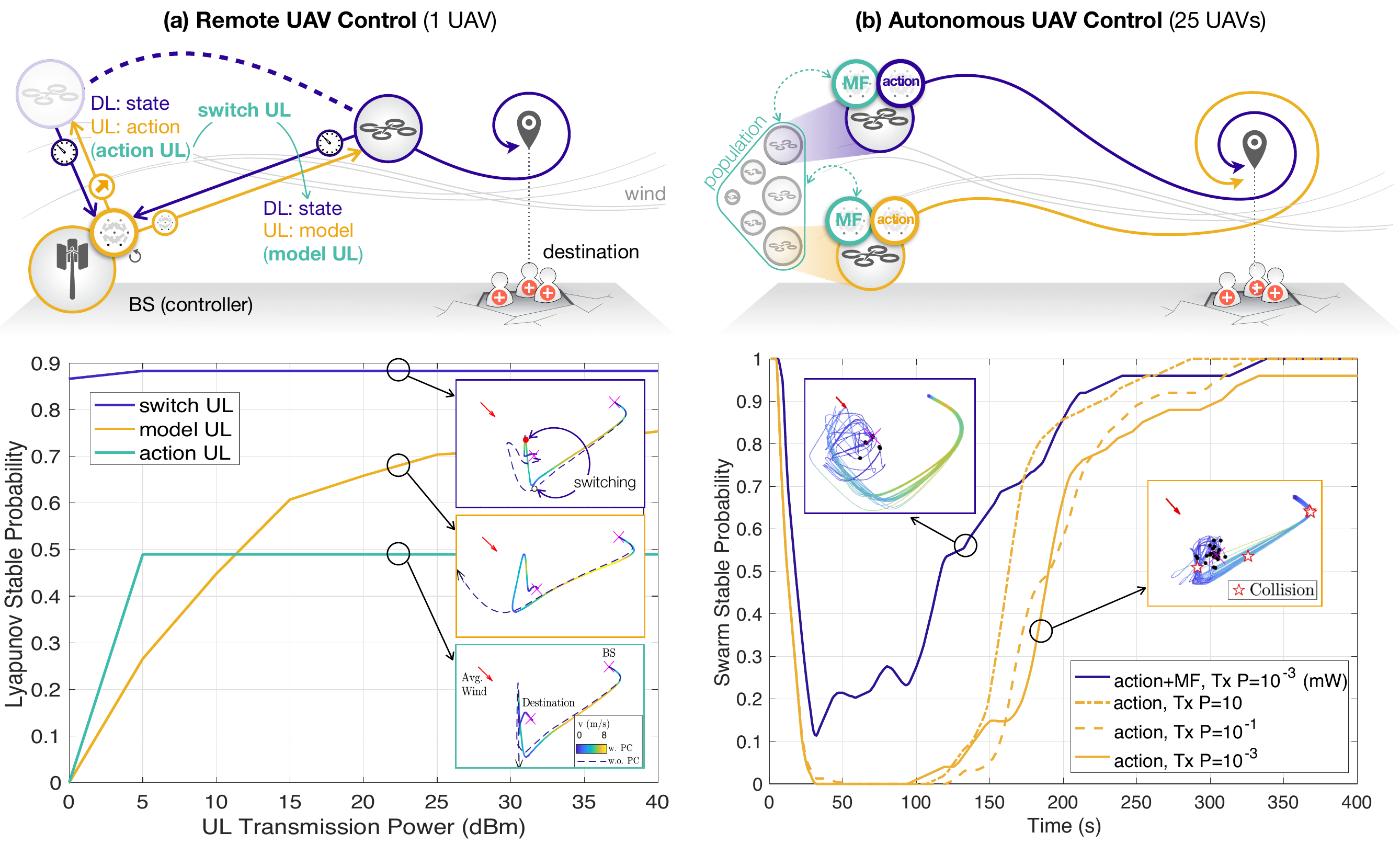} 
	\caption{\Gls{nn}-based single \gls{uav} control with its Lyapunov stability (destination arrival guarantee, left) and massive autonomous \gls{uav} control with their swarm stability (bounded maximum relative velocity guarantee, right).}
	\label{fig:use_case_control}
\end{figure*}

\subsection{Challenges and Opportunities}
Reaping the benefits of \gls{cococo} requires confronting several challenges, while opening novel research opportunities.

\vspace{5pt}\noindent\textbf{R7. Controller Connectivity.}\quad
A control system comprises sensors measuring states, controllers calculating commands based on the states, and actuators executing control commands. 
These three intertwined components are not always physically co-located, but connected over wireless links, resulting in missing and/or distorted state and command receptions. 
To alleviate this problem, by accounting for the sequential control operations, the controller's transmission power can be \emph{ramped up} if the preceding sensor's transmission delay is too long. 
By utilizing previous states and commands, future states and commands can also be \emph{inferred} via predictive \gls{urllc} (see \textbf{R3}). 
If future channel conditions are predicted to be poor, relocating the controller's functionality to its actuator allows to switch remote control to autonomous control.


\vspace{5pt}\noindent\textbf{8. Stable Control.}\quad
{When only communication reliability is considered, its associated control stability may be underestimated (e.g., useless communication attempts after collision) or overestimated (e.g., guaranteeing 9-nine communication reliability for achieving only 90\% stability). 
To correct this problem, control stability should first be clearly formalized.}
For a single agent control, when the output state is proportional to the input control, i.e., linear system,  system stability is determined by ensuring  bounded output for  any bounded input, i.e., \emph{\gls{bibo} stability}. 
In  non-linear systems,  stability can be examined through the lens of \emph{Lyapunov stability} that is ensured when the temporal derivative of the Lyapunov function decreases. 
For multiple-agent control, inter-agent collision avoidance can be described using \emph{swarm stability} that is satisfied when all agents' relative velocities converge to zero. 
For a chain of agents (e.g., vehicle platooning), alleviating chain fluctuations can be measured via \emph{string stability} that holds if any inter-agent spacing is bounded for a finite disturbance. 
Reconciling control stability constraints with communication reliability requirements under \textbf{R7} is a major challenge.

\vspace{5pt}\noindent\textbf{R9. Scalable Control.}\quad
In high-precision control applications, both control input and output dimensions are large incurring huge computing overhead. 
Moreover, in reality state dynamics  are often unknown due to  non-stationarity and external uncertainties, making traditional model-based control unfit for \gls{urllc} applications. 
\emph{\Gls{ml} based control} resolves both problems, in which an \gls{ml} model outputs an optimal control by directly feeding a state input. 
Another challenge comes from multiple interacting agents whose states are intertwined. Control decisions should thus be preceded by exchanging states, which may hinder scalability. 
\emph{\Gls{mf} game} framework elegantly detours this issue, in which each agent interacts only with the population's distribution that can be locally estimated.
Lastly, human intervention may interrupt machine operations due to their different perceptual characteristics (see \textbf{R3}), limiting scalability. 
In this respect, transferring human knowledge to machines via \emph{demonstrations} or \emph{human-machine \gls{fl}} is an emerging research direction.



\vspace{5pt}

%

%

%
\subsection{Use Cases}

%

	

%

\subsubsection{\glsentryshort{ml}-Aided Single \glsentryshort{uav} Remote Control}
Remote \gls{uav} control is an important use case highlighting the importance of \gls{cococo}. 
In order to control a remote \gls{uav} under random wind perturbations, the controller should download its state and upload the control decision to the \gls{uav} within a short time deadline. 
To meet this end-to-end control latency requirement, the effectiveness of uplink transmission power control and opportunistic controller relocation is studied as follows.

\vspace{5pt}\noindent\textbf{Scenario.}\quad
A single \gls{uav} is controlled by a ground \gls{bs} so as to reach a target destination. 
For each control cycle, the \gls{bs} \glspl{dl} the \gls{uav} state $s(t)$ (velocity and remaining distance) at time $t$, and runs an \gls{nn} (see \textbf{R9}) to compute its optimal action (acceleration) that is then uploaded to the \gls{uav}, until a time deadline, i.e., \gls{mad}. 
To meet the \gls{mad}, if the \gls{dl} latency is high, the \gls{ul} transmission power is ramped up. 
To cope with persistent remote control failures, when leaving a certain range, the \gls{ul} information is switched to the latest \gls{nn} model from its output control actions, enabling autonomous \gls{uav} control (see \textbf{R7}). 
Finally, in order not to pass by the destination, the \gls{nn} loss function is penalized when the Lyapunov stability $s(t)\text{d}s(t)/\text{d}t < 0$ is violated (see \textbf{R8}).

\vspace{5pt}\noindent\textbf{Results.}\quad
Action \gls{ul} payload sizes are smaller than model sizes. 
Hence, always uploading actions (\textsf{action \gls{ul}}) has faster control cycles with more state observations, yielding its better trained \gls{nn} model compared to always uploading NN models (\textsf{model~\gls{ul}}). 
However, even with  maximum \gls{ul} transmission power, \textsf{action~\gls{ul}} looses control of a faraway \gls{uav} that can be autonomously controlled under \textsf{model~\gls{ul}}. 
This trade-off is observed in Fig.~\ref{fig:use_case_control}(a). 
In comparison to these two baselines, by switching from \textsf{action~\gls{ul}} to \textsf{model~\gls{ul}}, the proposed control method (\textsf{switch~\gls{ul}}) ensures Lyapunov stability more frequently during the entire travel, while achieving shorter travel time.




\vspace{5pt}\subsubsection{\glsentryshort{ml} Aided Massive Autonomous \glsentryshort{uav} Control}
\gls{uav} swarms are critical in search and rescue missions, whereby forming a flock of \glspl{uav} can avoid inter-\gls{uav} collision, at the expense of exchanging instantaneous \gls{uav} states, hampering real-time control. 
\Gls{mf} game theoretic control alleviates the swarming communication overhead, enabling real-time control while avoiding collision. 
This is done by recasting the inter-\gls{uav} interactions as the interplay between a \gls{uav} and the population state distribution~\cite{park2018wireless}, as exemplified next.

\vspace{5pt}\noindent\textbf{Scenario.}\quad
There are 25 \glspl{uav} dispatched to a destination. 
Each \gls{uav} is autonomously controlled by locally running a pair of two \glspl{nn} (see \textbf{R9}), computing optimal control actions ({action \gls{nn}}) and population state distributions ({MF~\gls{nn}}), respectively. 
To avoid collision, the {action~\gls{nn}}'s loss function is penalized, when the maximum relative velocity is larger than a threshold, i.e., violating swarm stability $|v_\text{max} - v_\text{min}| > \varepsilon$ (see \textbf{R8}).

\vspace{5pt}\noindent\textbf{Results.}\quad
The convergence of the proposed control method (\textsf{action+MF}) is guaranteed, as long as the initial \gls{uav} states are exchanged.
Therefore, even with small transmission power (see \textbf{R7}), \textsf{action+MF} incurs no collision by achieving swarming faster than a benchmark scheme (\textsf{action}) running only action~\gls{nn} after exchanging instantaneous states, as observed by Fig.~\ref{fig:use_case_control}(b) .

\section{Conclusions}\label{sec:conclusions} 

This article outlined a detailed vision for the next generation of \gls{urllc}, coined \urllc{}. 
Breaking away from the reactive, \gls{rf} based, and communication centric 5G \gls{urllc},  \urllc{} is predictive, non-\gls{rf} aided, and weaves in  communication and control. 
This vision overcomes several key limitations of \gls{urllc}, namely extreme/rate event prediction, scalability, while building in new  diversity enhancements  with  minimal  overhead,  and  relaxing  latency and reliability requirements  based on the value of information. 
The intent of the \urllc{} vision is to spearhead beyond-5G/6G mission-critical applications (e.g., vision-based control, visuo-haptic \gls{vr}, autonomous/remote-controlled drone swarms, and other cyber-physical control applications). 
Going forward, \urllc{} can no longer be designed in a vacuum, but instead must leverage and build upon other domains and knowledge such as \gls{ml}, non-\gls{rf}, and control, while factoring in the cost of these  domains, notably with the era of data-driven decision-making and predictions.

\ifCLASSOPTIONcaptionsoff
  \newpage
\fi

\bibliographystyle{IEEEtran}
\bibliography{IEEEabrv}

\end{document}